\documentstyle[12pt, psfig,epsf]{article}
\begin{document}

\begin{center}
{\Large \bf A Feynman Diagram Analyser DIANA} \\

\vspace{4mm}
M.~Tentyukov\footnote{On leave of absence from
Joint Institute for Nuclear Research 141980 Dubna, Moscow Region, Russian Federation.
E-mail:
tentukov@physik.uni-bielefeld.de },
J.~Fleischer\footnote{E-mail: fleischer@physik.uni-bielefeld.de}\\
Fakult\"at f\"ur Physik, Universit\"at Bielefeld\\
D-33615 Bielefeld, Germany.

\begin{abstract}
A C-program
DIANA (DIagram ANAlyser) for the automatic Feynman diagram 
evaluation is presented. It consists of two parts:
the analyzer of diagrams and 
the interpreter of a special text manipulating language.
This language is used to create a source code 
for analytical or numerical evaluations and to keep the control of
the process in general.
\end{abstract}

\end{center}

\section{Introduction}
   Recent high precision experiments require, on the side of the theory,
high-precision calculations resulting in the evaluation of higher
loop dia\-grams in the Standard Model (SM).
For specific processes thousands of multiloop Feynman
dia\-grams do contribute, and it turns out to be impossible to perform 
these calculations by hand. This makes the request for automation a
high-priority task. 

Several different packages
have been developed with different areas of applicability (see a good 
review \cite{Steinh1}).
For example,
FEYNARTS / FEYNCALC \cite{FeynmArts} are MATHEMA\-TICA packages convenient
for various aspects of the calculation of radiative corrections in the SM.
There are several FORM packages for evaluating multiloop diagrams, like
MINCER \cite{MINCER}, and a package \cite{leo96} 
for the calculation of 3-loop bubble integrals with one non-zero mass.
Other packages for automation are
GRACE \cite{GRACE} and COMPHEP \cite{CompHep},
which partially perform full calculations, from the process definition 
to the cross-section values. 

A somewhat different approach is pursued  by
XLOOPS \cite{XLoops}. 
A graphical user interface
makes XLOOPS an `easy-to-handle' program package, but is mainly aimed 
to the evaluation of single diagrams.
To deal with thousands of diagrams, it is necessary to use 
special techniques like databases and special controlling programs.
In \cite{Vermaseren} for evalua\-ting more than 11000 diagrams
the special database-like program MINOS was developed.
It calls the relevant FORM programs, waits until they fi\-nished, picks 
up their results and repeats the process without any human interference.

The package GEFICOM \cite{GEFICOM}, developed for computation 
of higher order processes involving a large number of diagrams,
is based on cooperative usage of several software tools such as 
Mathematica, FORM, Fortran, etc.

It seems impossible to develop an universal package, which 
will be efficient for all tasks. It appears
absolutely necessary that various groups produce their own solutions
of handling the problem of automation:
various ways will be of different
efficiency, have different domains of applicability, and last but not
least, should eventually allow for completely independent checks of
the final results.
This point of view motivated us to seek our own
way of automatic evaluation of Feynman diagrams. 

Our first step was dedicated to the automation of 
the muons two-loop anomalous magnetic moment (AMM) 
${\frac{1}{2}(g-2)}_{\mu}$. For this purpose
the package TLAMM was developed \cite{TLAMM}.
The algorithm is implemented
as a FORM-based program package. For generating and automatically
evaluating any number of two-loop self-energy diagrams, a special
C-program has been written. This program creates the initial
FORM-expression for every diagram generated by 
QGRAF \cite{QGRAF}, executes the
corresponding subroutines and sums up the various contributions.
In the SM 1832 two-loop diagrams contribute in this case. 
The calculation of the bare diagrams is finished.

Our aim is to create some universal
software tool for piloting the process of generating the source
code in multi-loop order
for analytical or numerical evaluations and to keep the control of
the process in general. Based on this instrument, we can attempt to build 
a complete package performing the computation of any given process
in the framework of any concrete model. 

\section{Brief description}
For the project called DIANA (DIagram ANAlyser) for the
evalua\-tion of Feynman diagrams we have elaborated a special 
text manipulating language (TM).
The TM language is a very simple TeX-like language for creating
source code and organizing the interactive dialog.

The program reads QGRAF output. For each diagram it performs
the TM-program, producing input for further evaluation of the diagram.
Thus the program:

Reads QGRAF output and for each diagram it:
\begin{enumerate}
\item Determines the topology,
looking for it in the table of
all known topologies and distributes momenta.
If we do not yet know all needed topologies,
we may use the program
to determine missing topologies that occur in the process.
\item Creates an internal representation of the diagram 
in terms of vertices and propagators.
\item Executes the TM-program 
to insert explicit expressions for the vertices, propagators etc.
The TM-program produces input text for FORM ( or some other language),
 and executes the latter (optionally).
\end{enumerate}

Using the TM language, advanced users can develop further extensions,
e.g. including FORTRAN, to create
a postscript file for the  picture of the current diagram, etc.\\

\begin{figure}[ht]
\centerline{\epsfysize=130mm \epsfbox{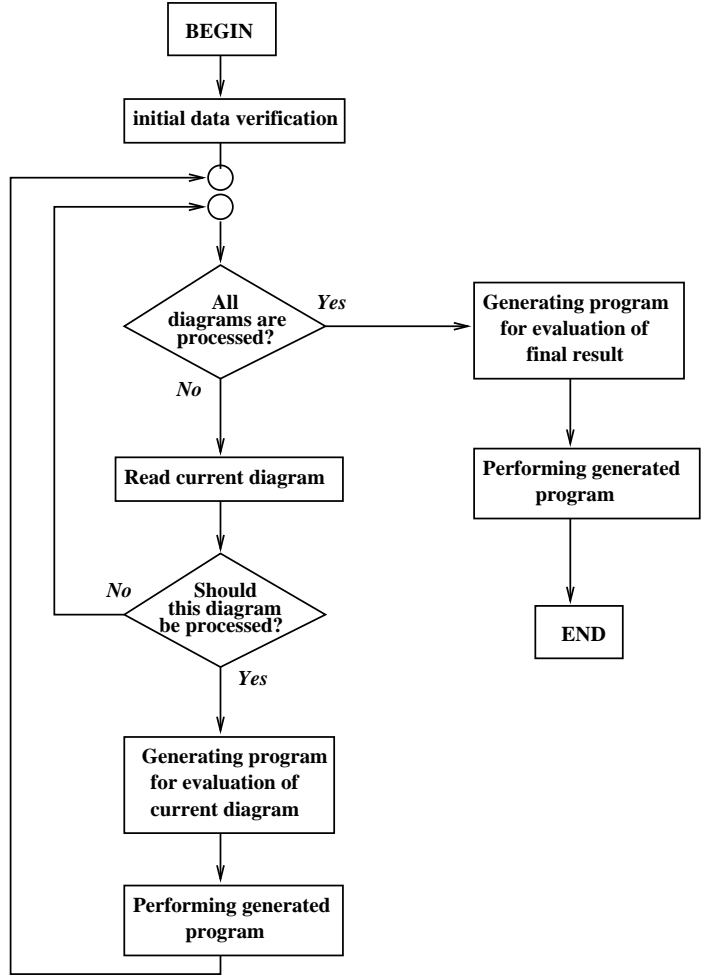}}
\caption{\label{dianam} Typical flowchart of Feynman diagram 
evaluations by DIANA.
    }
\end{figure}

The program operates as follows:
first of all, it reads its configuration 
file, which may be produced manually or by DIANA as well. This file contains:
\begin{enumerate}
\item The information about various settings (file names, numbers
of external particles, definition of key words, etc.)
\item Momenta distribution for each topology.
\item Description of the model (i.e., all particles, propagators and vertices).
\item TM-program.
\end{enumerate}
The TM-program is part of the configuration file. It starts with the
directive 
\small\begin{verbatim}
\begin translate
\end{verbatim}\normalsize

Then the program starts to read QGRAF output.
For each diagram it determines the topology,
assigns indices and creates the textual representation
of the diagram corresponding to the Feynman integrand. All defined data
(masses of particles, momenta on each lines, etc.) are stored
in internal tables, and may be called by TM-program operators.
At this point DIANA performs the TM-program.
After that it starts to work with the next diagram.

When all diagrams are processed, the program may perform the
TM-program a last time (optionally).
This may be used to do some final operations like summing up the
results.

Use of the TM language makes DIANA very flexible. It is easy to work out
various algorithms of diagram evaluation by specifying settings 
in the configuration file or even by a TM program. 

The typical flowchart may look like in Fig. \ref{dianam}. 

There is a possibility to use DIANA to perform the TM-program only,
without reading QGRAF output.
If one specifies in the configuration file
\small\begin{verbatim}
only interpret
\end{verbatim}\normalsize
then DIANA will not try to read QGRAF output, but immediately
enters the TM - program.

DIANA contains a powerful preprocessor. The user can create macros to hide 
complicated constructions.
Similar as LaTeX provides the possibility for non-specialists to
typeset high-quality texts using the TeX language, 
these macros permit DIANA to work at very high level. The user
can specify the model and the process, and DIANA will generate all necessary 
files. 

\section{Topologies and momenta}
\label{topologies}
\begin{figure}[ht]
\centerline{\epsfysize=40mm \epsfbox{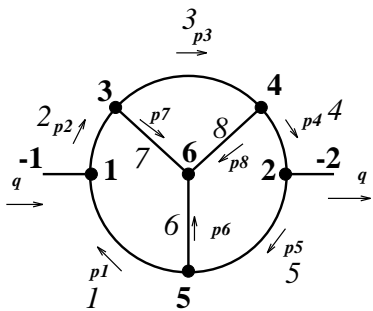}}
\caption{\label{topol1} Topology
(-2,2)(-1,1)(5,1)(1,3)(3,4)(4,2)(2,5)(5,6)(3,6)(4,6).
    }
\end{figure}

Topologies are represented in terms of
ordered pairs of numbers like (fromvertex, tovertex) (see Fig.~\ref{topol1}).
All external legs have negative numbers. These are supposed to begin with -1. 
First one attaches all the numbers for the ingoing particles, and then
for the outgoing ones.
 External legs must be connected with vertices of smallest possible 
identifying number. 
The number of an internal line corresponds to its position in the
chain of pairs and the direction from the first to the second number
in the pair:
$$
\begin{array}{ccccccccc}
\mbox{direction:}&5\to 1&1\to 3&3\to 4&4\to 2&2\to 5&5\to 6&3\to 6&4\to 6\\
(-2,2)(-1,1)&(5,1)&(1,3)&(3,4)&(4,2)&(2,5)&(5,6)&(3,6)&(4,6)\\
\mbox{number:}&1&2&3&4&5&6&7&8
\end{array}
$$
Knowing thus the topology we can assign momenta.
Their distribution according to Fig. \ref{topol1}, e.g., is added like 
\small\begin{verbatim}
topology A = (-2,2)(-1,1)(5,1)(1,3)(3,4)(4,2)(2,5)(5,6)(3,6)(4,6):
  p1,p2,p3,p4,p5,p6,p7,p8;
\end{verbatim}\normalsize
This fixes directions and values of all momenta on internal lines.
External lines must be known from the process definition. By
convention the external lines must start with the highest negative number.

All topologies have to be described in the configuration file. 
DIANA stores topologies in an internal table in some standard form.
After reading a new diagram DIANA defines its topology, reduces the 
topology
to the standard form and searches for the topology in the table.
If DIANA fails to find the topology, it will not produce 
the Feynman integrand for the diagram.

\section{TM language}

Similar to the TeX language, all lines without special escape - 
characters (``$\backslash$'') are simply typed to the output file. 
So, to type
``Hello, world!'' in the file ``hello'' we may write down the following
program:

\small\begin{verbatim}
\program
\setout(hello)
Hello, world!
\end{verbatim}\normalsize

The asterisk in the beginning of line is a comment:
\small\begin{verbatim}
* This is the comment!
\end{verbatim}\normalsize

Each word,
the first character of which is the escape character, will 
be considered as a command.
This feature makes this language very easy-to-use.
In appendix A there is an example of a simple TM-program and 
the result of its performance.

DIANA's preprocessor is run on the TM program before actual compilation.
It permits the user to do textual substitutions (with parameters),
to perform conditional compilation, etc.
For example, the following preprocessor directive
\small\begin{verbatim}
\DEF(macroname)
. . .
\ENDDEF
\end{verbatim}\normalsize
sets the macrodefinition. After this directive you can just write
\small\begin{verbatim}
\macroname()
\end{verbatim}\normalsize
to invoke the macro. You can use macros with arguments:
\small\begin{verbatim}
\macroname(a,b,c)
\end{verbatim}\normalsize
In the macro body the arguments are  available by the directive
\small\verb|\#(n)|, where ``n'' is the position of the argument.
The argument ``0'' is the macro name, all extra arguments are empty strings.

Example:
\small\begin{verbatim}
only interpret
\begin translate
\DEF(EXAMPLE) 0=\#(0),1=\#(1),2=\#(2),3=\#(3)\ENDDEF
\program
\offblanklines
\setout()
\EXAMPLE(a,b)
\end translate
\end{verbatim}\normalsize
The result is:
\small\begin{verbatim}
 0=EXAMPLE,1=a,2=b,3=
\end{verbatim}\normalsize

Using macros we are able to create a high level macrolanguage similar to 
LaTeX, which is the set of macrodefinitions under TeX. The basic idea 
is again similar to LaTeX, namely the use of {\it styles}. Instead of complicated 
TM programming one can use proper styles. The style file containing 
all macrodefinitions is just included in the beginning of the TM program.
We have several such styles oriented to the use of FORM. 

\section{Generating configuration files}

The style ``create.tml'' has the purpose to create the TM program. The
user has to prepare a file ``process.cnf'', which contains the model
and process specifications. Reading this file, DIANA will generate all
necessary other files. A typical example of such a file is the
following.
\small\begin{verbatim}
SET _processname = "zbb"
SET _qgrafname = "qgraf"
SET _syspath = "/home/U.Ser/diana/tml/"
system path = "/home/U.Ser/diana/tml/"
only interpret

\openlanguage(create.tml)
\Begin(program)

\Begin(model,gwsmassless.model)
\End(model)

\Begin(process)
ingoing Z(mu;p1);
outgoing b(;k1),B(;k2);
loops = 2;
\End(process)

\Begin(qgrafoptions)
options=onepi,nosnail;
\End(qgrafoptions)

\Begin(tmlprogram,form.prg)
\End(tmlprogram)

\indices(mu,mu1,mu2,mu3,mu4,mu5,mu6,mu7,
mu8,mu9,mu10,mu11,mu12,mu13,mu14,mu15,mu16,mu17,mu18,mu19,mu20,mu21,mu22,mu23)
\vectors(p1,p2,p3,p4,p5,p6,p,q,q1,q2,k1,k2)

\End(program)
\end{verbatim}\normalsize

This file contains all necessary information to create files for DIANA
to proceed the process $ Z \to b \bar b $ in the framework of the SM.
In the following we explain details of the above file.

\bigskip
Directives  of the type \small\verb|SET var = "val"|\normalsize{}
are used to set predefined preprocessor 
variables.  Thus, the directive 
\small\begin{verbatim}
SET _processname = "zbb"
\end{verbatim}\normalsize
just defines the name of the process. It will be used as filename extension
for all created files.
\small\begin{verbatim}
SET _qgrafname = "qgraf"
\end{verbatim}\normalsize
defines the command to invoke QGRAF.
The directives
\small\begin{verbatim}
SET _syspath = "/home/U.Ser/diana/tml/"
system path = "/home/U.Ser/diana/tml/"
\end{verbatim}\normalsize
are used to define the path to files containing TM macros and 
subroutines. The value of the preprocessor variable 
\small\verb|_syspath|\normalsize{} will be used by DIANA to 
define the line 
\small\begin{verbatim}
system path = "/home/U.Ser/diana/tml/"
\end{verbatim}\normalsize
in the created file ``settings.zbb''. In principle in ``process.cnf''
and ``settings.zbb'' these two paths may be different.

\bigskip 
The directive 
\small\begin{verbatim}
only interpret
\end{verbatim}\normalsize
tells DIANA not to read QGRAF output but just to perform the TM program.

\bigskip
The directive:
\small\begin{verbatim}
\openlanguage(create.tml)
\end{verbatim}\normalsize
is equivalent to the two directives
\small\begin{verbatim}
\begin translate
\include(create.tml)
\end{verbatim}\normalsize
The first one starts the TM language and the 
file \small\verb|create.tml|\normalsize{} contains 
all macro definitions and subroutines.
In particular, it contains the definition of the
macros \small\verb|\Begin()|\normalsize{} and \small\verb|\End()|. These
macros form an environment
depending on the first argument of the macro \small\verb|\Begin()|. 

\bigskip
The Body of the TM program is supposed to be placed between 
\small\begin{verbatim}
\Begin(program)
. . . 
\End(program)
\end{verbatim}\normalsize
Text after \small\verb|\End(program)|\normalsize{} will be ignored. 

\bigskip
The environment
\small\begin{verbatim}
\Begin(model,gwsmassless.model)
\End(model)
\end{verbatim}\normalsize
defines the model description. Instead of explicit definition,
in this example the file ``gwsmassless.model'' is used. This is a
file containing a simplified version of the SM.
It should be placed in 
the directory /home/U.Ser/diana/tml/ in order that DIANA can find it
in ``process.cnf''.

\bigskip
The process definition is clear from
\small\begin{verbatim}
\Begin(process)
ingoing Z(mu;p1);
outgoing b(;k1),B(;k2);
loops = 2;
\End(process)
\end{verbatim}\normalsize

\bigskip
The following environment defines the options passed to QGRAF \cite{QGRAF}:
\small\begin{verbatim}
\Begin(qgrafoptions)
options=onepi,nosnail;
\End(qgrafoptions)
\end{verbatim}\normalsize

\bigskip
The TM program for evaluating Feynman diagrams must be placed in the environment
\small\begin{verbatim}
\Begin(tmlprogram)
\End(tmlprogram)
\end{verbatim}\normalsize
Instead  one may specify the file containing the TM program as  second 
argument:
\small\begin{verbatim}
\Begin(tmlprogram,form.prg)
\end{verbatim}\normalsize
Here ``form.prg'' is the name of the file containing the standard TM program
for evaluation of Feynman diagrams by means of FORM.

\bigskip
The macros \small\verb|\indices()|\normalsize{} and
\small\verb|\vectors()|\normalsize{}
specify used 
indices and vectors. 

\section{The structure of the generated files}

Suppose the executable files ``diana'' and ``qgraf'' are available from
the system path and the files ``create.tml'', ``gwsmassless.model'',
``form.prg'' and ``process.cnf''
are situated in the directory ``/home/U.Ser/diana/tml/''
or in the current directory.

The user starts DIANA by means of the command
\small\begin{verbatim}
diana -c process.cnf
\end{verbatim}\normalsize
DIANA then generates several temporary files, invokes QGRAF, 
defines all topologies of the process and 
distributes momenta as follows:
the first line in each topology will carry momentum p1, the second p2, etc.
Three new files appear in the current directory:
\mbox{``config.zbb''}, ``qlist.zbb'' and ``settings.zbb''. 
The file ``config.zbb'' contains the TM program for the evaluation of the
Feynman diagrams of the specific process under consideration.
``settings.zbb'' 
contains various optional settings, topologies and momenta description,
the model and the process definition. 
It can be edited e.g. for the purpose of introducing proper integration
momenta for the various topologies by replacing the pi's.
The file ``qlist.zbb'' is the 
QGRAF output file.
To start the calculation of the Feynman diagrams the user finally 
enters the command
\small\begin{verbatim}
diana -c config.zbb
\end{verbatim}\normalsize

For the above example $ Z \to b \bar b $ 
the topologies in the file ``settings.zbb'' are:
\small\begin{verbatim}
topology top1_=
   (-3,3)(-2,2)(-1,1)(1,2)(1,3)(1,4)(2,4)(3,4):
      p1,p2,p3,p4,p5;
topology top2_=
   (-3,3)(-2,2)(-1,1)(1,3)(1,4)(1,4)(2,3)(2,4):
      p1,p2,p3,p4,p5;
topology top3_=
   (-3,3)(-2,2)(-1,1)(1,2)(1,4)(1,4)(2,3)(3,4):
      p1,p2,p3,p4,p5;
topology top4_=
   (-3,3)(-2,2)(-1,1)(1,4)(1,4)(2,3)(2,4)(3,4):
      p1,p2,p3,p4,p5;
topology top5_=
   (-3,3)(-2,2)(-1,1)(1,4)(1,5)(2,4)(2,5)(3,4)(3,5):
      p1,p2,p3,p4,p5,p6;
topology top6_=
   (-3,3)(-2,2)(-1,1)(1,2)(1,4)(2,5)(3,4)(3,5)(4,5):
      p1,p2,p3,p4,p5,p6;
topology top7_=
   (-3,3)(-2,2)(-1,1)(1,3)(1,4)(2,4)(2,5)(3,5)(4,5):
      p1,p2,p3,p4,p5,p6;
topology top8_=
   (-3,3)(-2,2)(-1,1)(1,4)(1,5)(2,3)(2,4)(3,5)(4,5):
      p1,p2,p3,p4,p5,p6;
topology top9_=
   (-3,3)(-2,2)(-1,1)(1,2)(1,3)(2,4)(3,5)(4,5)(4,5):
      p1,p2,p3,p4,p5,p6;
topology top10_=
   (-3,3)(-2,2)(-1,1)(1,2)(1,4)(2,3)(3,5)(4,5)(4,5):
      p1,p2,p3,p4,p5,p6;
topology top11_=
   (-3,3)(-2,2)(-1,1)(1,3)(1,4)(2,3)(2,5)(4,5)(4,5):
      p1,p2,p3,p4,p5,p6;
\end{verbatim}\normalsize
The user must edit this file to set the desired integration momenta 
and the internal numeration of
lines according to the rules described in Sect. \ref{topologies}. 
In the near future the pictorial representation of the topologies will
be implemented.

\bigskip
As next we give a detailed description of
the file ``config.zbb''. It contains the directive to include the file 
``settings.zbb'', several settings for various options and
a skeleton  of the FORM program in terms of the TM language. The
following is a listing of the file. The whole TM-program is divided into
different ``sections''. Each section will be performed only under
proper conditions.

\begin{quote}
\small\begin{verbatim}
* This file is automatically generated by DIANA for the process zbb.
\include(settings.zbb)

*Remove the following to avoid the generation of the protocol file:
log file = log.zbb 

* Remove the following line to avoid debug information:
debug on

extra call

\openlanguage(specmode.tml)
\Begin(program,routines.rtn)

\section(regular)

\Begin(initialization)
  \export(callform,\ask(\(Call form?(Y/N))))
\End(initialization)

\Begin(output,\askfilename())
   **** d\counter()
   **** (diagram \currentdiagramnumber())
   \blankline()
   * Set here your defines!
   \Begin(foreach,i,1,\numberofinternallines())
      #define mm\i() "\mass(\i())"
      \blankline()
   \End(foreach)
   #define LINE "\numberofinternallines()"
   #define FERMIONLINE "\maxfcount()"
   #define TOPOLOGY "\topologyid()"
   \blankline()
   functions \functions();
   commuting \commuting();
   l Rq =\integrand();
   \blankline()
   * Here should be your FORM program!
   drop Rq;
   g dia\counter()=Rq;
   .store
   save dia\counter().sto;
   .end
\End(output)
\Begin(special,"\import(callform)"eq"Y")
   \execute(\(form -l )!.!)
\End(special)

\section(epilog)

   \Begin(output)
      #do j = 1,\counter()
        load dia'j'.sto;
        .store
      #enddo
      g Rq =
      #do j = 1,\counter()
         + dia'j'
      #enddo
      ;
      .sort
      print;
      .end 
   \End(output)

   \Begin(special,"\import(callform)"eq"Y")
      \execute(\(form -l )!.!)
   \End(special)

\End(program)

\end{verbatim}\normalsize
\end{quote}
The setting \small\verb|log file = log.zbb|\normalsize{} induces the 
creation of the protocol file 
``log.zbb''.

\bigskip
If the user sets \small\verb|debug on|, DIANA will produce a detailed 
record of the translation process. Some information necessary for 
possible run-time diagnostics will be stored as well.

\bigskip
The setting \small\verb|extra call|\normalsize{} forces DIANA to perform
the TM-program once again after all diagrams are processed. In this
case only
the section \small\verb|epilog|\normalsize{} will be called (see below).

In the next command the TM-program uses the macro - style ``specmode.tml'':
\small\begin{verbatim}
\openlanguage(specmode.tml)
\end{verbatim}\normalsize
This style defines the ``sections'' mechanism: a TM-program
may consist of several ``sections'', each of them being performed
only under certain conditions. So the generic structure of the
TM-program looks like follows:
\small\begin{verbatim}
\Begin(program)
\section(sectionname)
   (content...)
. . . 
\End(program)
\end{verbatim}\normalsize

\bigskip
The main section is \small\verb|\section(regular)|. It is performed
to create FORM input for each diagram.

The section 
\small\begin{verbatim}
\section(epilog)
\end{verbatim}\normalsize
will be performed during ``extra call'' after all diagrams have been processed.
In the above example this section is used to sum up the final results.

The whole TM-program is put into the environment
\small\begin{verbatim}
\Begin(program,routines.rtn)
\End(program)
\end{verbatim}\normalsize
It contains the argument ``routines.rtn''. This is the name of a file
containing various  TM-functions and macros, which are loaded in this
manner, e.g.\\
\small\verb|\functions()|\normalsize{} -- outputs a comma - separated list of all
non-commuting functions occuring in the current diagram and\\
\small\verb|\commuting()|\normalsize{} -- outputs the list of all
commuting functions.

In our example the global variable ``callform'' is defined by means of
the function \small\verb|\ask()|\normalsize{}. This
function outputs the argument to the screen waits for the user to reply
``y'' or ``n'' and returns ``Y'' or ``N'', respectively,
i.e. ``callform'' then takes the value ``Y'' or ``N''. This action is
placed in the environment
\small\begin{verbatim}
\Begin(initialization)
 . . .
\End(initialization)
\end{verbatim}\normalsize
and thus is performed only once.

The section ``\small\verb|regular|\normalsize'' contains the skeleton
of the FORM program.
Output, i.e. the FORM input, inside the environment 
\small\begin{verbatim}
\Begin(output,filename)
 . . .
\End(output)
\end{verbatim}\normalsize
is directed  into the file ``filename''. In the above example, instead
of an explicit file name,  we use the macro 
\small\verb|\askfilename()|\normalsize{} defined in
the style file \small\verb|specmode.tml|\normalsize. 
This macro asks the filename from the user. 
For each diagram a filename of the type``\small\verb|d.frm|'' will be expanded to 
``\small\verb|d#.frm|'', where \small\verb|#|\normalsize{} is the
number of 
the current diagram. Thus, for diagram number 15 the file ``\small\verb|d15.frm|''
will be created. Note that this expansion is performed by the
environment and
not by the macro \small\verb|\askfilename()|.

The macro \small\verb|\askfilename()|\normalsize{} asks the file name
only once in the case when the file name is not defined yet. 
It is impossible to change the file name once entered. 

The macro \small\verb|\counter()|\normalsize{} counts the 
processed diagrams. It can differ from the 
diagram number (produced by QGRAF for the set of all diagrams)
returned by the operator \small\verb|\currentdiagramnumber()|.

The operator \small\verb|\blankline()|\normalsize{} 
returns a blank line. It must be
used to obtain a blank line in the produced output because all blank
lines in the TM-program are  suppressed by default.

There is the environment \small\verb|foreach|\normalsize{} providing a 
``cycle with parameter''.
The content of the environment 
\small\begin{verbatim}
\Begin(foreach,i,1,10)
\End(foreach)
\end{verbatim}\normalsize
will be repeated 10  times. Each time the macro \small\verb|\i()|\normalsize{} expands to
the digit counter, i.e 1,2,...10.

The built-in operator \small\verb|\numberofinternallines()|\normalsize
returns for each diagram the number
of internal lines; the built-in operator
\small\verb|\maxfcount()|\normalsize{} 
returns the number of connected fermion lines in the current diagram. 

The macro \small\verb|\integrand()|\normalsize{} is expanded in a sequence of 
TM-operators providing the Feynman integrand for the current diagram.

There is the environment \small\verb|special|\normalsize. It has the form 
\small\begin{verbatim}
\Begin(special,<condition>)
 . . .
\End(special)
\end{verbatim}\normalsize

The content of this environment will be performed only if 
\small\verb|<condition>|\normalsize{} evaluates to \small\verb|True|.

In the folowing example this environment is used to call FORM:
\small\begin{verbatim} 
\Begin(special,"\import(callform)"eq"Y")
   \execute(\(form -l )!.!)
\End(special)
\end{verbatim}\normalsize
The macro \small\verb|\execute()|\normalsize{} expands  the symbols 
\small\verb|!.!|\normalsize{} to the full file name (e.g.,
\small\verb|d15.frm|\normalsize).
The expression \small\verb|!.|\normalsize{} is expanded
to the file name without extension, and \small\verb|.!|\normalsize{} 
is expanded to the extension.
The macro \small\verb|\execute()|\normalsize{} calls the built-in TM-operator
\small\verb|\system()|\normalsize{} to execute FORM. 
Thus, for the diagram number 15,
e.g., the following command will be fulfilled:
\small\begin{verbatim}
form -l d15.frm
\end{verbatim}\normalsize
Here we assume that the user has entered the filename ``d.frm''.

The section ``\small\verb|epilog|'' uses the environment ``\small\verb|output|''
without a second argument. The second argument is unnecessary
because the file name is already defined.

In the ``extracall'' the environment ``\small\verb|output|''  uses the
filename exactly the user has specified it, answering the question
from the macro \small\verb|\askfilename()|, i.e. ``d.frm'' in this case.

\section{Generating the FORM input and executing FORM}

DIANA has many command line options. In particular, the user can
specify the numbers of the diagrams to be evaluated. 
The command 
\small\begin{verbatim}
diana -c config.zbb -b 451 -e 453
\end{verbatim}\normalsize
tells DIANA to process diagrams number 451, 452 and 453. After executing
this command 
four new files appear in the current directory. If the user
has entered the file name ``d.frm'' these files are:
``d451.frm'', ``d452.frm'', ``d453.frm'' and ``d.frm''.
The content of the file ``d.frm'' looks like follows:
\small\begin{verbatim}
      #do j = 1,3
        load dia'j'.sto;
        .store
      #enddo
      g Rq =
      #do j = 1,3
         + dia'j'
      #enddo
      ;
      .sort
      print;
      .end 
\end{verbatim}\normalsize
It can be used by FORM to sum up the calculated diagrams.

The file ``d452.frm'' looks like follows:
\small\begin{verbatim}
**** d2
**** (diagram 452)

* Set here your defines!
#define mm1 "mmt"

#define mm2 "mmt"

#define mm3 "0"

#define mm4 "0"

#define mm5 "0"

#define mm6 "mmt"

#define LINE "6"
#define FERMIONLINE "2"
#define TOPOLOGY "top8_"


functions F,FF;
commuting VV;
l Rq =
       (-1)*F(2,1,mu1,1,0,1)*(-i_)*em*Qd*FF(3,1,-p3,0)*i_*
       F(3,1,mu3,1,0,1)*(-i_)*em*Qd*F(1,2,mu,GVu,-GAu,1)*(+i_)*em/2/s/c*
       FF(1,2,-p1,mt)*i_*F(4,2,mu2,1,0,1)*(-i_)*em*Qu*FF(6,2,-p6,mt)*i_*
       F(5,2,mu4,1,0,1)*(-i_)*em*Qu*FF(2,2,+p2,mt)*i_*
       VV(4,mu1,mu2,+p4,0)*i_*VV(5,mu3,mu4,+p5,0)*i_;

* Here should be your FORM program!
.end

\end{verbatim}\normalsize

From the beginning of this file we can see that the macro 
\small\verb|\counter()|\normalsize{} produces
the order number
of the processed diagram, i.e. 2, while the built-in TM operator
\small\verb|\currentdiagramnumber()|\normalsize{} produces the number
of the current diagram
according to the QGRAF file ``qlist.zbb'' i.e. 452.

\begin{figure}[ht]
\small\begin{verbatim}
. . .
*--#[ d452:
*
     -1
    *vx(B(-2),b(1),A(2))
    *vx(B(1),b(-4),A(3))
    *vx(T(5),t(4),Z(-1))
    *vx(T(4),t(6),A(2))
    *vx(T(6),t(5),A(3))
*
*--#] d452: 
. . .
\end{verbatim}\normalsize
\vskip -47mm
\hspace*{7cm}{\epsfysize=40mm \epsfbox{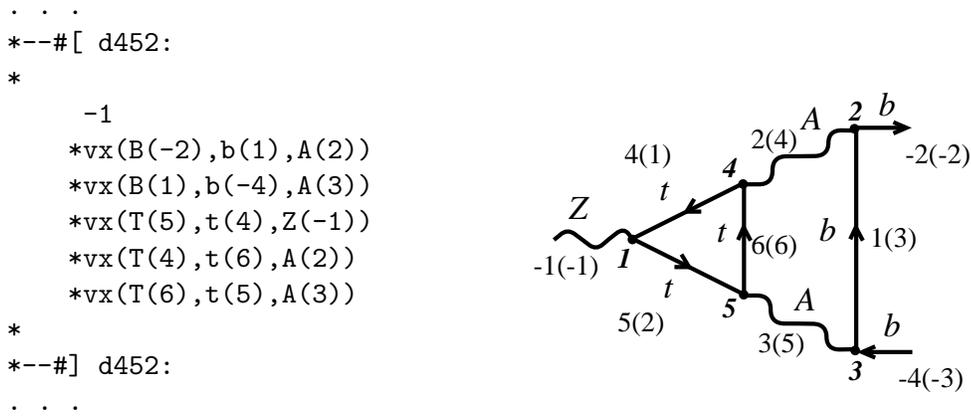}}
\vskip 5mm
\caption{\label{d452}To the left there is the  QGRAF output for the
diagram ``number 452''. {\tt t} and {\tt T} stand for
t and anit-t quarks, {\tt b} and {\tt B} stand for b and anit-b
quarks, {\tt A} is the photon, {\tt Z} is the Z boson.
External legs are numerated by negative numbers and {\tt vx} stands for the
vertex. 
The corresponding graph is shown on the right.
The lines are numbered corresponding to QGRAF output. The numbers in
brackets correspond to the numbering defined by ``topology 8'' in the file
``settings.zbb''. Bold italic digits numerate vertices according to
the topology 8.}
\end{figure}

In Fig. \ref{d452} the QGRAF output  for the diagram number 452 is
shown together with the corresponding graph. 
We can see that in this case there are two fermion lines,
one corresponding to the external $b-\bar b$, while the second appears
as closed fermion loop, the $t$ quark triangle. That is why the
\small\verb|#define|\normalsize{}  ``FERMIONLINE'' is equal to ``2''.

In order to understand the Feynman integrand structure in the above example
we
consider the simplified SM in use, i.e. the ``gwsmassless.model''. In 
our example is contained in the file
``settings.zbb''. The following notation is used:\\
{\tt em}  -- electromagnetic coupling ($em > 0$);\\
{\tt s}   -- $\sin(\theta_W)$;\\
{\tt c}   -- $\cos(\theta_W)$;\\
{\tt mt, mW, mZ, mH} -- masses;\\
{\tt mmt, mmW, mmZ, mmH} -- masses squared;\\
{\tt Qe, Qu, Qd} -- electric charges in units of the proton charge,
i.e. -1, 1/3, -2/3;\\
{\tt GAe, GAnu, GAu, GAd} -- axial couplings;\\
{\tt GVe, GVnu, GVu, GVd} -- vector couplings;

We have two (c)functions for propagators: {\tt FF} and {\tt VV}.\\
Vector propagator:\\
  {\tt VV(}
\begin{quote}
1.~ number of line\\
2.~index of the first particle\\
3.~index of the second particle\\
4.~type (0 -- photon, 1 -- {\tt Z}, 2 -- {\tt W}, 3--gluon)
\end{quote}
  {\tt )}\\

  Fermion propagator:\\
  {\tt FF(}
\begin{quote}
1.~number of line\\
2.~fermion line number\\
3.~ mass
\end{quote}
  {\tt )}\\

  For vertices we have one commuting function {\tt V} and one function {\tt F}.
\begin{itemize}
  \item The first argument always is the number of the vertex.
  \item The last argument shows the type of the vertex:
    0 -- scalar, 1 -- vector, 2 -- tensor, 3 -- VVV, 4 -- VVVV
  \item The function {\tt F} carries as second argument the fermion line
    number.
\end{itemize}

For example, the fermion-vector-fermion vertex is:\\
    {\tt F(}
\begin{quote}
1.~number of line\\
2.~fermion line number \\
3.~index\\
4.~{\tt GV}  (vector coupling)\\
5.~{\tt GA}  (axial coupling)\\
6.~1  (the type of vertex
\end{quote}
  {\tt )}

Note that the ``gwsmassless.model'' assumes a zero {\tt b} quark mass (see 
\small\verb|#define mm3 "0"|\normalsize).

\section{Conclusion}

  Higher order loop calculations and/or multileg ones in one-loop
order in the SM of electroweak interactions require
the calculation of many complicated Feynman diagrams, which is due 
to the large particle spectrum of the model. The number of diagrams
can exeed many thousands. Therefore the evaluation of such large
number of diagrams needs to be automated in almost all respects.
First of all an automation of the diagram generation is needed.
Since FORM, in our opinion, is the most suitable language for such 
kind of calculations,
we found it of urgent need to write a program for this purpose which
directly produces ``FORM input'', i.e. produces an algebraic representation
of Feynman diagrams in terms of their momentum representation, written
in FORM format. This is what the presented program DIANA is able to do 
at present and we believe it will have wide applications. For an application
of DIANA for the process $Z \to b \bar{b}$ see (\cite{Zbb}). 

   In further steps techniques of handling the numerators and
evaluation of scalar Feynman diagrams must be included. These methods
partially exist already (\cite{TLAMM,LME}).

\vskip 10mm 
\noindent
{\large \bf Acknowledgments}
\vskip 10mm
\noindent 
We are grateful to M.~Kalmykov and O.~Veretin
for helpful discussions.
M.T was supported by Bundesministerium f\"ur Forschung 
und Technologie under PH/05-7BI92P 9. and in part by RFBR $\#$98-02-16923.
\vskip 10mm 
\noindent
\appendix{\large \bf Appendix A. Example of a simple TM-program}
\vskip 10mm 
\noindent
In this appendix we consider a ``pure'' TM-program {\em without} 
macrocoding. This should help to understand some basis of the TM
language.

Let us suppose that we use FORM as formulae manipulating language.

The user types his FORM program and has the possibility to insert 
in the same line TM-operators as well.
For example, the typical part of a TM program looks like follows:
\small\begin{verbatim}
\program
\setout(d\currentdiagramnumber().frm)
#define dia "\currentdiagramnumber()"
#define TYPE "\type()"
#define COLOR "\color()"
#define LINES "\numberofinternallines()"
\masses()
#include def.h
l  R=\integrand();
#call feynmanrules{}
#call projection{}
#call reducing{'TYPE'}
#call table{'TYPE'}
#call colorfactor{'COLOR'}
.sort
g dia'dia' = R;
drop R;
.store
save dia'dia'.sto;
.end
\setout(null)
\system(\(form -l )d\currentdiagramnumber().frm)
\end{verbatim}\normalsize

Some of the TM - commands are just TM-operators while some are functions 
(returning a value) written in 
the TM-language itself.

Only one type of data exists in the TM language. This is a string.
All operators and functions return only one text string. 
They may have one or more arguments,
or may have no argument at all. Arguments are separated by commas.
Number of arguments is fixed for each operator or function.
The default value returned by any function is an empty string. 

 The following operators are built-in TM operators:\\
\small\verb|\numberofinternallines()|\normalsize{} -- returns the
number of internal lines.\\
\small\verb|\currentdiagramnumber()|\normalsize{} -- returns the order number of current
diagram.\\
\small\verb|\setout(filename)|\normalsize{} -- redirects output to the file
``filename''. There is the special file ``null''; output redirected to
this file never appears. If an empty string is used as argument of this
operator, the output will appear at the terminal.
The above operator always returns an empty string.\\
\small\verb|\system(cmd)|\normalsize{} -- calls the operating system to perform the
command ``cmd''. It returns some integer number for the system status
of this command.\\

An argument of any operator may contains an arbitrary number of
other operators.  All spaces in the operator arguments are suppressed.

There is the special ``quotating operator'' without name, it just returns
its argument without any change:
\small\verb|\(any text)|\normalsize{} returns the string ``\small\verb|any text|''.
In the above example it is used in the argument of the operator 
\small\verb|\system()|\normalsize{} to keep spaces. 

The functions
\small\begin{verbatim}
\masses(), \type(), \color() ,\integrand()
\end{verbatim}\normalsize
are TM-language functions. Their code should be placed somewhere
before the main
program (the latter  starts with the operator \small\verb|\program|).
As an example we present the function ``masses'':
\small\begin{verbatim}
\function masses;\-\let(i,0)
\+\do
#define m\inc(i,1) "\mass(\get(i))"
\while "\numcmp(\get(i),\numberofinternallines())" eq "<" loop
\end
\end{verbatim}\normalsize
This function just outputs a string like 
\small\begin{verbatim}
#define m1 "mmH"
\end{verbatim}\normalsize
for each internal line. 
The built-in operator \small\verb|\mass(1)|\normalsize{} returns the
mass of the particle at
line 1.

Some string may be stored in a variable. The operator \small\verb|\let(var,val)|
creates (or resets) variable \small\verb|var|\normalsize{} 
and gives it the value \small\verb|val|\normalsize{}.
The operator \small\verb|\get(var)|\normalsize{} returns \small\verb|val|:
\small\begin{verbatim}
\program
\let(who,world)
\setout()
Hello, \get(who)!
\end{verbatim}\normalsize
Note, the operator \small\verb|\let(var, val)|\normalsize{} returns 
\small\verb|val|, not an empty string!

There are several built-in operators for manipulating integer
numbers. 
In the above example of the 
function \small\verb|\masses()|\normalsize{} the operator 
\small\verb|\inc(i,1)|increments value of \small\verb|i|\normalsize{} 
by \small\verb|1|\normalsize{}  
and returns the new value. The operator 
\small\verb|\numcmp(a,b)|\normalsize{} returns ``\small\verb|<|'', 
``\small\verb|>|'' or
``\small\verb|=|'', depending on the numerical 
values of \small\verb|a|\normalsize{} and
\small\verb|b|.

Variables defined by the operator \small\verb|\let()|\normalsize{} are
 local w.r.t.
TM-functions. The user can define global variables by means the
operator \small\verb|\export(var,val)|. Global variables exist during
processing all diagrams. They can be obtained by the operator
\small\verb|\import(val)|.

Unlike built-in operators, the user - defined functions can 
produce some output to the current output file. 
The function \small\verb|\masses()|\normalsize{}, e.g., returns an
empty string, but it produces several lines placing them into
the output file. That is why we use derictives
\small\verb|\-|\normalsize{} and \small\verb|\+|: the former just
suppresses
 any output while
the latter again permits output.

Functions can
return arbitrary strings by means of the operator \small\verb|\return(text)|.

There are several kinds of ``controlling constructions''. 
They are used to change the stream of the control.
These are the well - known ``if'', ``do'', etc. In the function
\small\verb|\masses()|\normalsize{} you
can see how the post-condition cycle is used.

This TM-program will generate the FORM input for each diagram.
For example, the corresponding part of 
the FORM program generated for diagram number 15 will be placed into the
file ``d15.frm'' and looks like follows:
\small\begin{verbatim}
#define dia "15"
#define TYPE "4"
#define COLOR "3"
#define LINES "4"
#define m1 "mmH"
#define m2 "mmW"
#define m3 "mmW"
#define m4 "mmH"
#include def.h
l  R=
       1*V(1,mu1,mu,2)*(-i_)*em^2/2/s*V(2,0)*(-i_)*1/4*em^2/s^2*mmH/mmW*
       V(3,mu2,+q4-(+q3),1)*(-i_)*em/2/s*SS(1,0)*i_*VV(2,mu1,mu2,+q2,2)*i_*
       SS(3,2)*i_*SS(4,0)*i_;
#call feynmanrules{}
#call projection{}
#call reducing{'TYPE'}
#call table{'TYPE'}
#call colorfactor{'COLOR'}
.sort
drop R;
g dia'dia' = R;
.store
save dia'dia'.sto;
.end
\end{verbatim}\normalsize

Then this program will be performed by FORM 
by means of the operator
\small\begin{verbatim}
\system(\(form -l )d\currentdiagramnumber().frm)
\end{verbatim}\normalsize
In this case the operator performs the command
\small\begin{verbatim}
form -l d15.frm
\end{verbatim}\normalsize{}

\end{document}